\documentstyle[prd,aps,epsfig,floats,axodraw]{revtex}
\begin{document}

\draft
\input epsf \renewcommand{\topfraction}{0.8}
\newcommand{\beq}{\begin{equation}}
\newcommand{\eeq}{\end{equation}}
\newcommand{\pbar}{\not{\!\partial}}
\newcommand{\dbar}{\not{\!{\!D}}}
\def\lsim{\:\raisebox{-0.75ex}{$\stackrel{\textstyle<}{\sim}$}\:}
\def\gsim{\:\raisebox{-0.75ex}{$\stackrel{\textstyle>}{\sim}$}\:}
\twocolumn[\hsize\textwidth\columnwidth\hsize\csname@twocolumnfalse\endcsname
\title{Sleptogenesis}
\author{Rouzbeh Allahverdi$^{a}$, Bhaskar Dutta$^{b}$ and Anupam Mazumdar
$^c$}
\address{$^a$TRIUMF, 4004 Wesbrook Mall, Vancouver, BC, V6T
2A3, Canada. \\
$^{b}$Department of Physics, University of Regina, Regina, SK, S4S 0A2,
Canada. \\
$^{c}$CHEP, McGill University, 3600 Uiversity Road, Montr\'eal, QC,
H3A 2T8, Canada.}
\maketitle
\begin{abstract}
We propose that the observed baryon asymmetry of the Universe can naturally 
arise from a net 
asymmetry generated in the right-handed sneutrino sector at fairly low reheat 
temperatures. The initial asymmetry in the sneutrino sector is produced from 
the decay 
of the inflaton, and is subsequently transferred into the Standard Model
(s)lepton doublet via three-body decay of the sneutrino. Our scenario relies 
on two main assumptions: a considerable branching ratio for the inflaton 
decay to the right-handed (s)neutrinos, and Majorana masses which are 
generated by the Higgs mechanism. The marked 
feature of this scenario is that the lepton asymmetry is decoupled from the 
neutrino Dirac Yukawa couplings. We exhibit
that our scenario can be embedded within minimal models which seek
the origin of a tiny mass for neutrinos.
\end{abstract}


\vskip2pc]

\section{Introduction}

The consistency of the abundance of the light elements synthesized
during the big bang nucleosynthesis (BBN) requires that the baryon
asymmetry of the Universe (BAU) parameterized as
$\eta_{\rm B}=(n_{\rm B}-n_{\bar{\rm B}})/s$, with $s$ being the entropy
density, and $n_{B}$ the number density of the baryons, be in the
range $(0.3-0.9)\times 10^{-10}$~\cite{Olive:1999ij}. This asymmetry can be
produced from a baryon symmetric Universe provided three conditions
are simultaneously met; $B$ and/or $L$-violation, $C$-and $CP$-violation,
and departure from thermal equilibrium~\cite{Sakharov:dj}. Any produced
asymmetry will however be washed away by the standard model (SM)
$B+L$-violating sphaleron transitions which are active from temperatures
$10^{12}$~GeV down to $100$~GeV~\cite{Kuzmin:1985mm}, if $B-L=0$. Therefore an
asymmetry in $B-L$, which is subsequently reprocessed by sphalerons, is
generally sought in order to yield the net baryon asymmetry given by
$B=a(B-L)$. Here $a$ is a model-dependent parameter; in case of
the SM, $a=28/79$, while in the minimal supersymmetric standard model
(MSSM), $a=32/92$~\cite{Khlebnikov:sr}.

An attractive mechanism for producing $B-L$ asymmetry is from the
decay of the heavy right-handed (RH) Majorana neutrinos
\cite{Fukugita:1990gb}. Since the RH neutrinos are the SM singlets,
a Majorana mass $M_N$, which violates lepton number, is compatible
with all of its symmetries, and hence can be arbitrarily large
beyond the electroweak scale. This provides a natural explanation
for the light neutrinos via the see-saw mechanism~\cite{seesaw}.

The lepton asymmetry can be generated from the interference between the
tree-level and the one-loop diagrams in an out-of-equilibrium decay of
the RH neutrinos in the early Universe, provided $CP$-violating
phases exist in the neutrino Yukawa couplings. The asymmetry thus
obtained will be partially converted into the baryon asymmetry via sphaleron
effects. This is the standard lore for producing lepton asymmetry from
on-shell RH neutrinos, commonly known as leptogenesis
\cite{Fukugita:1990gb,Luty:un,plumacher98}. This can be accomplished in 
different ways.

In thermal leptogenesis scenario, RH neutrinos come into equilibrium
with the primordial thermal bath through Yukawa interactions. The
decay of the lightest RH neutrino easily satisfies the out-of-equilibrium
condition by virtue of having a sufficiently small Yukawa
coupling~\cite{plumacher98}. In a model-independent analysis in
Ref.~\cite{Buchmuller:2002rq}, the authors have parameterized thermal
leptogenesis by four parameters; the $CP$ asymmetry, the heavy
RH neutrino mass, the effective light neutrino mass, and the quadratic
mean of the light neutrino masses. The final result was that an acceptable
lepton asymmetry could be generated with
$T_{\rm R} \sim M_{1}={\cal O}(10^{10})$~GeV, and
$\sum_{i}m_{\nu,i}< \sqrt{3}$~eV.

This is marginally compatible with the upper
bound on $T_{\rm R}$ allowed from thermal gravitino production in 
supersymmteric models \cite{ellis}. Gravitinos with a mass 
${\cal O}(\rm TeV)$ decay long after nucleosynthesis and their 
decay products can change abundance of the light elements synthesized 
during BBN. For $100~{\rm GeV}\leq m_{3/2}\leq 1~{\rm TeV}$, a 
successful nucleosynthesis requires $n_{3/2}/s \leq (10^{-14}-10^{-12})$, 
which translates into $T_{\rm R}\leq (10^{8}-10^{10})~{\rm GeV}$ 
\cite{ellis,subir}. The possibility of non-thermal gravitino production 
\cite{maroto} does not give rise to any threat as described in 
\cite{rouzbeh,nps}. It was also suggested that gravitinos
can also be produced directly from the inflaton decay~\cite{nop}, and 
in the decay of heavy stable neutral particles~\cite{allahverdi01}, but 
the yielded bounds will not be severe.

An interesting alternative is non-thermal leptogenesis. This could
happen in many ways. The simplest possibility is to produce on-shell RH 
neutrinos, with a considerable branching ratio, in inflaton decay~
\cite{kumekawa94}. It is also possible to produce heavy RH neutrinos (even 
heavier than the inflaton) via preheating \cite{giudice99}. However, 
non-thermal leptogenesis is rather model dependent. For example, just 
fermionic preheating is plagued by the fact that the running coupling 
of the inflaton to the fermions can easily give rise to correction in the 
inflaton mass, which leads to the instabilities during the inflaton 
oscillations, described as in Ref.~\cite{enqvist02a}. The inflaton   
condensate fragments as a result of that and forms interesting 
solitons.

In supersymmetric models there are additional options as one can also 
excite sneutrinos~\cite{cdo}. In fact sneutrinos are produced more abundantly 
than neutrinos during preheating~\cite{bdps}. Another possibility is creating 
a 
condensate of sneutrinos which yields the right asymmetry through 
its decay~\cite{yanagida}, or via Affleck-Dine 
mechanism~\cite{berezhiani01}.

Recently it has been noticed that successful leptogenesis does not
require on-shell RH (s)neutrinos~\cite{bento,am}. A minimal model was 
proposed in Ref.~\cite{am}, where the lepton asymmetry is directly generated
from the inflaton decay into the Higgs and leptons via off-shell RH
(s)neutrinos. This model naturally results in a sufficiently low
reheat temperature, and yields desirable baryon asymmetry for a
rather wide range of inflationary scale, neither invoking  
preheating in a particular model, nor, any unnaturally suppressed 
couplings.

In this paper, we propose a completely new scenario for leptogenesis, 
called sleptogenesis\footnote{Baryogenesis with scalar fields has also been 
studied in Ref.~\cite{mrs}, though in a different context.}. We show that an 
asymmetry between sneutrinos 
and anti-sneutrinos can be generated, through a phase mismatch between 
the inflaton coupling to the RH (s)neutrinos and the Majorana masses, in 
inflaton decay. 
Note that the RH neutrino and anti-neutrino are indistinguishable due 
to the Majorana nature of neutrinos. After the (s)neutrinos decay, 
the SM (s)leptons carry the produced asymmetry which will be partially 
reprocessed to the baryon asymmetry. This scenario can emerge quite naturally 
provided the branching ratio for the inflaton decay to the RH (s)neutrinos is 
considerable, and there exists new Higgs field(s) generating  
the Majorana masses. 
The first assumption is rather common in non-thermal scenarios of 
leptogenesis, while the latter is necessary in models where the RH 
(s)neutrinos are gauge non-singlet under some new physics. The main feature 
of our scenario 
is replacing the dependence of the generated asymmetry on the neutrino Dirac 
Yukawa couplings with that on the Majorana Yukawa couplings. 
As a consequence, it is in principle possible to accommodate low-scale 
leptogenesis~\cite{hambye} with an appropriate choice of model parameters.
The minimal extension of MSSM that can accommodate the above mentioned 
Majorana sector is  
$SU(3)_c \times SU(2)_L \times U(1)_{I_{3R}} \times U(1)_{B-L}$. In this 
scenario, the RH (s)neutrino
masses  arise at the scale where $U(1)_{I_{3R}} \times U(1)_{B-L} \rightarrow 
U(1)_Y$.
The branching ratios of lepton flavor violating decay modes e.g. 
$\tau\rightarrow\mu\gamma$, $\mu\rightarrow
e\gamma$ will be able to discern these  models in the near future.

\section{The scenario}

We begin by considering a simple model in a supersymmetric set up. The
relevant part of the superpotential is given by
\begin{eqnarray}
\label{superpot}
W &\supset & \frac{1}{2} m_\phi \Phi^2 + \frac{1}{2} m_\sigma \Sigma^2 +
\frac{1}{2}{\bf y} \Phi {\bf N}^2 + \frac{1}{2} {\bf g} \Sigma {\bf N}^2 \,
\nonumber \\
&+ & {\bf h} {\bf N} {\bf H}_u {\bf L} + h_t {\bf H}_u {\bf Q}_3 {\bf t}^{c}\,.
\end{eqnarray}
Here $\Phi$ is a gauge singlet superfield which comprises the inflaton 
$\phi$ and its superpartner (inflatino) with mass $m_\phi$, and ${\bf N}$ 
is the superfield comprising the RH neutrino $N$ and sneutrino 
$\widetilde N$. While $\Sigma$ comprises the scalar field $\sigma$ 
which generates Majorana mass for $N$ through its VEV, denoted as 
$\sigma_0$, and its fermionic partner $\widetilde \sigma$. As we will 
describe later, in realistic particle physics models ${\bf N}$ and $\Sigma$ 
are charged under some gauge group (as a matter of fact, one needs to 
introduce another superfield ${\bar \Sigma}$ for anomaly cancellation).
Since the inflaton is assumed to be a gauge singlet,
its coupling to RH (s)neutrinos actually arises at the non-renormalizable 
level, and hence is small 
${\bf y}\sim {\cal O}(m_{\phi}/M_{\rm P})$ (we use the reduced Planck mass
$M_{\rm P}\sim 2.4\times 10^{18}$~GeV). This coupling will be responsible for 
decay of the inflaton to $N$ and ${\widetilde N}$ and, subsequently, 
reheating the Universe.

Finally, ${\bf H}_u$, ${\bf L}$, ${\bf Q}_3$, and ${\bf t}^{c}$ are the
multiplets containing the Higgs which gives mass to the top quark,
the left-handed lepton doublet, the third generation quark doublet
and the RH top anti-quark, along with their superpartners, respectively.
We have omitted all indices on ${\bf N}$, and lepton doublets.
Note that ${\bf y}$ and ${\bf g}$ are symmetric matrices. For 
simplicity, we assume that they can be diagonalized in the same 
basis, and hence only their diagonal elements $y_i$ and $g_i$ are 
relevant.

We also assume that $m_\sigma \geq 10 m_\phi$. This implies that
the dynamics of $\sigma$ is frozen during and after inflation, and hence 
ensures a simpler dynamics by the virtue that all of the 
energy density is carried by $\phi$. However, 
the mass of the RH (s)neutrinos $M_i$ (at least one of them) 
is taken to be smaller than $m_\phi$, so that the inflaton decay
to $N_i$ and ${\widetilde N}_i$ will reheat the Universe\footnote{Note that 
we have neglected another 
coupling of the form $f \Phi {\Sigma}^{2}$ , even though it can 
arise at the renormalizable level in realistis models, and hence need not 
be very small. The reason is that in the limit $m_\sigma \gg m_\phi$, such a 
coupling can only affect the inflaton decay by inducing 
$\phi \rightarrow {\widetilde N} 
{\widetilde N} {\widetilde N} {\widetilde N}$, via off-shell $\sigma$ and 
${\widetilde \sigma}$, and $\phi \rightarrow 
{\widetilde N} {\widetilde N}$ decay modes at the tree-level and one-loop 
level, 
respectively. The effective coupling for these modes will be $f (g 
m_{\phi}/M)^{2}$ and $fy$, respectivley, and, moreover, their decay rate 
is suppressed by four-body phase space factor and one-loop factor, 
respectively. Thus the inflaton predominantly decays via 
coupling $y$, and a coupling between $\Phi$ and $\Sigma$ will have no 
bearings on our results.}

An important point is that the interference between the tree-level
and one-loop contributions to the decay process 
$\phi\rightarrow{\widetilde N}{\widetilde N}$ results in an excess, 
or deficit, of $\widetilde N$ over ${\bar{\widetilde N}}$, provided 
a relative phase exists between $g_i$ and $y_i$. This happens in 
exactly the same fashion as $N$ decay generates a lepton asymmetry 
in the standard leptogenesis scenario~\cite{Fukugita:1990gb}.

Note that it is meaningless to talk of any asymmetry between
$N$ and $\bar N$, since there is no distinction between particle
and anti-particle for a Majorana fermion. To put it another way,
the mass term $M_N NN$, which violates the lepton number, makes
particle and anti-particle indistinguishable. On the other hand,
the supersymmteric mass term $M^{2}_{N}|\widetilde N|^2$ for the 
sneutrino does not violate the lepton number.

In most of the realisitc models of inflation only the real component of the 
inflaton has a VEV. Then it can be shown from Eq.~(\ref{superpot}) that $\phi 
\rightarrow N N$ and $\phi \rightarrow {\widetilde N} {\widetilde N}$ decays 
occur at the same rate, and the total decay rate
is given by 
\beq 
\label{decay} 
\Gamma_{\rm d} \simeq {1\over 8 \pi} \sum_{i} y^{2}_{i} m_\phi\,. 
\eeq 
Note that $\Delta L = 2$ in $\phi \rightarrow {\widetilde N}{\widetilde N}$ 
decay. By taking into account of the one-loop self-energy and vertex 
diagrams, shown in Fig.~(1), we find that\footnote{There are also 
contributions from supersymmetry breaking terms to these diagrams which will 
be suppressed as $m_{3/2}/m_\sigma$.} 
\begin{center}
\SetScale{0.6} \SetOffset(40,25)
\begin{picture}(200,100)(0,0)
\DashArrowLine(0,50)(50,50){5} \Text(0,25)[l]{$\phi$}
\Vertex(50,50){3}
\ArrowArc(75,50)(25,0,180) \Text(45,55)[t]{$N$}
\ArrowArcn(75,50)(25,0,180) \Text(45,5)[b]{$N$}
\Vertex(100,50){3}
\Vertex(150,50){3}
\DashArrowLine(100,50)(150,50){5} \Text(75,40)[t]{$\sigma$}
\DashArrowLine(150,50)(200,100){5} \Text(130,65)[r]{$\widetilde N$}
\DashArrowLine(150,50)(200,0){5} \Text(130,0)[r]{$\widetilde N$}              
\end{picture}
\vspace*{-13mm}
\end{center}

\vspace*{3mm}

\begin{center}
\SetScale{0.6} \SetOffset(40,25)
\begin{picture}(200,100)(0,0)
\DashArrowLine(0,50)(75,50){5} \Text(0,25)[l]{$\phi$}
\Vertex(75,50){3}
\ArrowLine(125,100)(75,50) \Text(55,55)[t]{$N$}
\ArrowLine(125,0)(75,50) \Text(55,5)[b]{$N$}
\Vertex(125,100){3}
\Vertex(125,0){3}
\ArrowLine(125,100)(125,0) \Text(90,30)[r]{$\widetilde \sigma$}
\ArrowLine(125,0)(125,100)
\DashArrowLine(125,0)(200,0){5} \Text(135,65)[r]{$\widetilde N$}
\DashArrowLine(125,100)(200,100){5} \Text(135,5)[r]{$\widetilde N$}           
\end{picture}
\vspace*{-13mm}
\end{center}

\vspace*{5mm}

\noindent
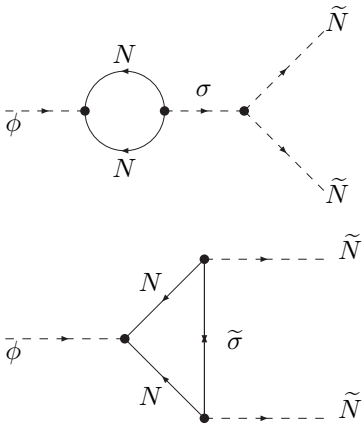
{\bf Fig. 1:}~One-loop self-energy and vertex diagrams resulting in an 
asymmetry between $\widetilde N$ and $\bar {\widetilde N}$. 
\vspace*{4mm}
\newline
\beq 
\label{asymmetry}
{n_{{\widetilde N}_{i}}-n_{{\bar {\widetilde N}}_{i}}\over n_\phi}= -{1
\over 8 \pi} {{\rm Im} [{({\bf y} {\bf g}^{\dagger})}_{ii}]^2 \over
\sum_{i} {({\bf y}{\bf y}^{\dagger})}_{ii}}~ f \left({m^{2}_{\sigma} \over
m^{2}_{\phi}}\right)\,, 
\eeq 
where 
\beq 
\label{x} 
f(x) = {\sqrt{x} \over 2} \left[{2\over x-1} + {\rm ln} \left(1 + 
{1\over x}\right) \right]\,. 
\eeq 
These diagrams are similar to those
in leptogenesis via ${\widetilde N}$ decay~\cite{cdo} (with 
proper replacements). The expression for the asymmetry 
parameter therefore has exactly the same structure as in the standard 
leptogenesis~\cite{one-loop}. There are slight differences though 
between the two cases. Here only half of the 
inflatons decay to RH sneutrinos, and $\phi$ decay to $N$ does not lead to 
any asymmetry. On the other hand, lepton number is violated by two units in 
$\phi \rightarrow {\widetilde N} {\widetilde N}$ decay. Finally, a factor of 
$1/2$ arises in our case since identical particles appear in the loop. Note 
that
in the limit $m_\sigma \geq 10 m_\phi$, we simply have $f \simeq 3m_\phi/
2m_\sigma$\footnote{In the limit $m_{\phi}=m_{\sigma}$ the 
perturbative results in Eqs. (\ref{asymmetry}),(\ref{x}) break down. In this 
case one has to actually take into account of the finite decay width of 
$\phi$ and $\sigma$. 
This has been done for the standard lepotogenesis with degenerate Majorana 
(s)neutrinos, and it is shown that no asymmetry will be yielded, as expected, 
in the $x=1$ limit \cite{pilaftsis}.}.  

The created asymmetry is then transferred into the SM (s)leptons via
${\widetilde N}_i$ decay. There are two two-body decay channels read 
from Eq.~(\ref{superpot}): 
${\widetilde N}_i \rightarrow {\bar L}_i{\bar{\widetilde H}}$ and 
${\widetilde N}_i \rightarrow {\widetilde L}_i H$, which have the same 
rate. Here $h_i$ denotes diagonal elements 
of the neutrino Yukawa matrix ${\bf h}$ and, for simplicity, we assume 
that non-diagonal elements can be neglected. Since the two-body decays 
produce the same number of anti-leptons as leptons, no net lepton 
asymmetry will be yielded.

However, there exists a term 
$h_ih_t{\widetilde N}_{i}{\widetilde L}{\bar{\widetilde Q}}_{3}
{\bar{\widetilde t}^{c}}$ in the scalar potential which results in the 
three-body 
decay ${\widetilde N}_{i}\rightarrow{\bar{\widetilde L}}{\widetilde Q}_3
{\widetilde t}^{c}$. This channel is responsible for transferring the 
asymmetry 
into the SM (s)leptons, though with suppression by a factor $\simeq 3/32\pi^2$ 
(note that $h_t\approx 1$). The $1/32\pi^{2}$ is the ratio of phase space 
factors for three-body decay to the total decay rate, and note that 
$\widetilde N$ 
decays to all three colours of squarks. In addition, we also have the usual 
dilution due to the entropy release from reheating by a factor of 
$T_{\rm R}/m_\phi$, where $T_{\rm R}$ denotes the reheat temperature. A 
thermal bath of the SM particles(and their superpartners) is typically formed 
right after $\widetilde N$ and $N$ decay (for details on thermalization, see 
Ref.~\cite{thermal}), and hence $T_{\rm R}$ is determined by the details of 
these decays.

Here we assume that all ${\widetilde N}_i$ (and $N_i$) decay very rapidly
right after they have been produced. This will simplify the calculations while
preserving the essence of our scenario. It will be the case
if $\Gamma_i\geq\Gamma_{\rm d}$, where $\Gamma_i$ is the decay
rate of ${\widetilde N}_{i}$ (and, by virtue of supersymmetry, $N_i$). The
(s)neutrinos, with mass $M_i$, initially having an energy $\simeq m_\phi/2$,
and hence their decay rate (at the time of production) is given by
\beq \label{Ndecay}
\Gamma_i \simeq {h^{2}_{i}M^{2}_{i} \over 2 \pi m_\phi}.
\eeq
Note that the decay rate at the ${\widetilde N}_{i}$ rest frame is $h^{2}_{i} 
M_{i}/4$, and the time-dilation factor will be $2 m_{\phi}/M_{i}$.

The requirement that the (s)neutrinos decay when $H\simeq\Gamma_{\rm d}$,
translates into the condition $4 h^{2}_{i}M^{2}_{i} \geq y^{2}m^{2}_{\phi}$,
where $y^2=\sum_{i} y^{2}_{i}$. In the minimal see-saw model the limit on the 
light neutrino masses, with the current 
cosmological and laboratory bounds on the absolute neutrino masses taken into 
account, translates to   
\beq 
\label{mlimit} 
{h^{2}_{i}\langle H^{0}_{u} \rangle^2 \over M_i} \leq 10^{-9}~{\rm GeV}\,, 
\eeq
where $\langle H^{0}_{u} \rangle \simeq 174$ GeV is the Higgs VEV. 
Since $M_i < m_\phi$, the instant (s)neutrino decay requires that 
$y^2< 10^{-14}(m_{\phi}/1~{\rm GeV})$. This results in a tiny $y$, 
which also fulfills the requirement from the model building point of view. A 
small coupling $y$ also ensures a sufficiently low $T_{\rm R}$.

After putting all the pieces together, including the reprocessing by
sphalerons and dilution from reheating, we obtain
\beq
\label{baryon}
\eta_{\rm B} \simeq {9 \over 64 \pi^2} \cdot {1 \over 8 \pi} {\sum_{i}
y^{2}_{i} g^{2}_{i} \over y^2} {T_{\rm R} \over m_\sigma},
\eeq
where
\beq \label{reheat}
T_{\rm R} \simeq  {g^{1/4}_{*} \over 3} (y^2 M_{\rm P} m_\phi)^{1/2}.
\eeq
Here $g_*$ is the number of relativistic degrees of freedom ($g_* \simeq 200$ 
in the MSSM when $T_{\rm R} > 1$ TeV). Note that $n_{\phi} \simeq g_{*} 
T^{4}_{\rm R}/3 m_{\phi}$, while $s \simeq g_{*}T^{3}_{\rm R}/{\pi}^{2}$.

Let us denote ${\widetilde N}_1$ as the sneutrino which makes the largest
contribution to the asymmetry. Then Eq.~(\ref{baryon})
implies that it
has the largest combination $yg$, but not necessarily the largest $y$, or,
$g$. Note that the inflaton mainly decays into the (s)neutrino with the
largest $y$, while the heaviest (s)neutrino has the largest coupling $g$
(see Eq.~(\ref{superpot})). The maximum asymmetry is yielded when 
$y_1>y_2,y_3$. For $y^2 \simeq y^{2}_{1}$ the expression in
Eq.~(\ref{baryon}) is further simplified to
\beq
\label{simple}
\eta_{\rm B} \simeq {g^{2}_{1} \over 2^9 \pi} {T_{\rm R} \over
m_\sigma}\,.
\eeq
Therefore a successful leptogenesis requires that
\beq
\label{limit1}
g^{2}_{1} {T_{\rm R} \over m_\sigma} \gsim 5 \times 10^{-8}\,.
\eeq
A couple of important comments are in order now. The preservation of 
the lepton number by the sneutrino mass term has been a key point in 
our scenario. This is true for the sneutrino supersymmetric mass 
derived from the superpotential. However, supersymmetry must be broken 
in any realistic model and this inevitably introduces soft breaking 
terms. The soft breaking mass term $m^{2}_{3/2}|{\widetilde N}|^2$, 
with $m_{3/2}$ being the gravitino mass, also preserves the lepton number. 
On the other hand, the $A$-term associated with the Majorana mass term, 
which has the form $a m_{3/2} M_N {\widetilde N}{\widetilde N}+{\rm h.c.}$, 
breaks the lepton number in the sneutrino sector. This term will cause an 
oscillation between the sneutrino and anti-sneutrino, similar to the 
neutrino flavor oscillations, with a frequency $am_{3/2}$. In consequence, 
any asymmetry between $\widetilde N$ and $\bar {\widetilde N}$ only survives 
for a time $\leq (a m_{3/2})^{-1}$, while being washed out by 
${\widetilde N}-{\bar{\widetilde N}}$ oscillations at longer time 
scales. Therefore the success of our proposed scenario requires 
that ${\widetilde N}_1$ decay early enough, i.e. $\Gamma_1 \geq a m_{3/2}$.

The value of $m_{3/2}$ depends on the mechanism for communicating
supersymmetry breaking to the observable sector. In gravity-mediated models
$m_{3/2} \simeq 100~{\rm GeV}-1$ TeV, while in gauge-mediated models
substantially smaller values $m_{3/2} \simeq 1$ KeV are possible. The
situation then depends on the exact value of $a$, which is determined by
the structure of K\"ahler potential. For minimal K\"ahler terms one
typically has $a \simeq {\cal O}(1)$, while $a \approx 0$ can be obtained
in non-minimal cases. Let us focus on the former case, as it will clearly
result in a more stringent bound. Then it is required that
\beq 
\label{Mlimit1}
{h^{2}_{1} \over 8 \pi} M_1 \geq 10^{2}~(10^{-6})~{\rm GeV}\,, 
\eeq
in gravity (gauge)-mediated models in order to preserve the lepton asymmetry. 
By taking into account of the see-saw constraint in Eq.~(\ref{mlimit}), we
obtain the absolute lower bound 
\beq 
\label{Mlimit2} M_1 \geq
10^{8.5}~({10}^{4.5})~{\rm GeV}\,, 
\eeq 
on the mass of the RH neutrino with
largest contribution to the asymmetry. Note that the above bound is only 
meant for the minimal K\"ahler structure and can be significantly weakened 
for non-minimal kinetic terms.

\section{Wash-out of the generated asymmetry}

We now turn our attention to various interactions which can wash out
the produced asymmetry. First, let us briefly recount thermal
history of the Universe in our scenario. The inflaton mainly decays
into the ${\bf N}_1$ multiplet when $H \simeq \Gamma_{\rm d}$, and
a lepton asymmetry is generated in the decay to the sneutrino component
${\widetilde N}_1$. Then ${\widetilde N}_1$, as well as other (s)neutrinos, 
decays promptly and we obtain a thermal bath consisting of the SM
degrees of freedom (and their superpartners) with temperature $T_{\rm R}$
estimated in Eq.~(\ref{reheat}).

The first lepton-number violating interaction is the $N$ and 
${\widetilde N}$-mediated scattering of leptons and Higgs (also their 
superpartners) in a
thermal bath. These scattering have been considered in detail in 
the standard leptogenesis scenario \cite{fy,Luty:un,plumacher98}. As
an illustration; a sample scattering of this type will be inefficient, 
only if
\beq
\label{erase1}
\Gamma_{\not{L}} \simeq {h^4 \over 16 \pi^3} {T^{3}_{\rm R} \over
M^{2}_{N}} < g^{1/2}_{*} {T^{2}_{\rm R} \over M_P}\,.
\eeq
Note that there exists a large number of such scattering, especially in
the MSSM~\cite{plumacher98}.

By using the relationship in Eq.~(\ref{mlimit}), we obtain the constraint 
on reheat temperature which will avoid erasure of the lepton asymmetry. 
This bound turns out to be smaller than the gravitino overproduction 
bound $T_{\rm R} < 10^{10}$~GeV.

There are also other lepton number violating interactions, namely
the $\widetilde \sigma$ and $\sigma$-mdeiated ${\widetilde N}_{1} 
{\widetilde N}_{1}$ and ${\widetilde N}_{1} N_1$
scatterings, shown in 
Fig.~(2)\footnote{Note that $\widetilde\phi$ and $\phi$-mediated scatterings 
can be neglected due to the smallness of the inflaton coupling to 
$\widetilde N_{1}$ and $N_{1}$.}. These processes can erase the
lepton asymmetry carried by ${\widetilde N}_1$ before it decays, 
provided they occur at a higher rate. Note that the number density of $N_1$ 
and ${\widetilde N}_1$ is $\simeq g_* T^{4}_{\rm R}/3 m_\phi$. This will 
result in
\beq \label{erase2}
\Gamma_{{\widetilde N}_1 N_1} \simeq {C g^{4}_{1} \over 24 \pi^3} 
{g_* T^{4}_{\rm R} \over m_\phi m^{2}_{\sigma}},
\eeq
and
\beq \label{erase3}
\Gamma_{{\widetilde N}_1 {\widetilde N}_1} \simeq {C g^{4}_{1} \over 12 \pi^3} 
{g_* T^{4}_{\rm R} \over m_\phi m^{2}_{\sigma}},
\eeq
\begin{center}
\SetScale{0.6} \SetOffset(40,35)
\begin{picture}(150,100)(0,0)
\DashArrowLine(0,100)(75,100){5} \Text(0,50)[l]{$\widetilde N$}
\Vertex(75,100){3}
\DashArrowLine(0,0)(75,0){5} \Text(0,10)[l]{$\widetilde N$}
\Vertex(75,0){3}
\ArrowLine(150,100)(75,100) \Text(90,50)[r]{$N$}
\ArrowLine(150,0)(75,0) \Text(90,10)[r]{$N$}
\ArrowLine(75,100)(75,0) \Text(55,30)[r]{$\widetilde \sigma$}
\ArrowLine(75,0)(75,100)
\end{picture}
\vspace*{-13mm}
\end{center}

\vspace*{3mm}

\begin{center}
\SetScale{0.6} \SetOffset(40,35)
\begin{picture}(150,100)(0,0)
\DashArrowLine(0,100)(75,100){5} \Text(0,50)[l]{$\widetilde N$}
\Vertex(75,100){3}
\ArrowLine(0,0)(75,0) \Text(0,10)[l]{$N$}
\Vertex(75,0){3}
\DashArrowLine(150,100)(75,100){5} \Text(90,50)[r]{$\widetilde N$}
\ArrowLine(150,0)(75,0) \Text(90,10)[r]{$N$}
\DashArrowLine(75,100)(75,0){5} \Text(55,30)[r]{$\sigma$}
\end{picture}
\vspace*{-13mm}
\end{center}

\vspace*{2mm}

\noindent
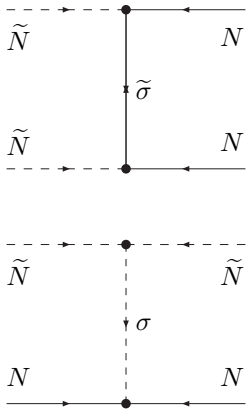
{\bf Fig. 2:}~Processes violating lepton number in the sneutrino sector. 
\vspace*{4mm}
\newline
where $C$ is a multiplicity factor representing different 
contribuitions to the same process, and $C/{\pi}^2 \sim {\cal O}(1)$. Also 
recall the 
decay rate $\Gamma_1 
\simeq {h^{2}_{1}M^{2}_{1}/4 \pi m_\phi}$ 
for ${\widetilde N}_1$. With the help of Eqs.~(\ref{mlimit},\ref{limit1}) 
we find that these processes will be inefficient, provided 
\beq \label{limit2} 
\left({T_{\rm R} \over
1~{\rm GeV}}\right)^2 < 10^{-2} \left({M_1 \over 1~{\rm
GeV}}\right)^3\,. 
\eeq
Note that $T_{\rm R} < M_1$ for a perturbative decay of $N_1$ (for details 
see Ref.~\cite{thermal}). Therefore this bound is easily satisfied as long 
as $M_{1}> 100$ GeV. In conclusion, the only non-trivial constraint in our 
scenario will be that of generating sufficient asymmetry, given in 
Eq.~(\ref{limit1}).

So far we have only considered the
${\widetilde N}_{1} {\widetilde N}_{1}$ and ${\widetilde N}_{1} N_1$ 
scatterings. On the other hand, 
${\widetilde N}_{1} {\widetilde N}_{1} \rightarrow N_{i} N_{i}$ and
${\widetilde N}_{1} N_1 \rightarrow {\widetilde N}_{i} N_{i}$
annihilaitions can also happen through diagrams in Fig.~(2). The rate for 
such processes is 
$\propto g^{2}_{1}g^{2}_{i}$, which will be larger than the one considered 
above, provided $N_1$ is not the heaviest RH neutrino (note that 
$M_i \propto g_i$). However, as we shall see shortly, successful 
baryogenesis requires that $M_1$ not be much smaller than $m_\phi$. This 
implies that $M_1$ is not very different from the largest $M_i < m_{\phi}$, 
and hence 
the rate for various processes represented by diagranms in Fig.~(2) are in 
general comparable. 
Moreover, Eq.~(\ref{erase2}) will indeed give 
the largest rate if $N_1$ is the heaviest RH neutrino.

\section{Model Parameters}

We can now estimate the range of parameters within which our scenario can
accommodate a successful baryogenesis. As an example;
$m_\sigma \gsim 10 m_\phi$, which guarantees that $\sigma$ does not play
any dynamical role in post-inflationary era, while from
Eq.~(\ref{limit2}); $M_{1} \gsim 10 T_{\rm R}$ guarantees the survival
of generated asymmetry. Then the observed baryon asymmetry can be
obtained provided
\beq
\label{range}
g^{2}_{1} {M_1 \over m_\phi} \gsim 5 \times 10^{-6}\,.
\eeq
For $g_1 \gsim 10^{-2}$,  this would require that $M_1$ be (at least) an 
order of
magnitude smaller than $m_\phi$. This is at par with the standard
non-thermal leptogenesis where (s)neutrinos are produced perturbatively. Note 
that a smaller $M_1/m_\phi$ is allowed as $g_1$ increases.

It is important to notice that, contrary to the standard leptogenesis 
scenario, sufficient asymmetry can be obtained with much smaller values 
of $M_1$. In fact, it is evident from Eq.~(\ref{limit1}) that $\eta_{\rm B}$ 
only depends on the ratio $M_1/m_\phi$. Therefore, as advertised earlier, our 
scenario can accommodate low scale leptogenesis 
without making unntaural assumptions (e.g. having highly degenertae Majorana 
neutrinos, Ref.~\cite{hambye}). This is a consequence of generatring the 
lepton asymmetry directly in the inflaton decay, and hence decoupling it 
from the neutrino Dirac Yukawas. One should nevertheless keep in mind 
the lower bound on $M_1$, from Eq.~(\ref{Mlimit2}), which arises for the 
minimal K\"ahler potential. However, this has an entirely different origin, 
namely to avoid the earsure of the asymmetry by $\widetilde N-\bar 
{\widetilde N}$ oscillations induced by soft supersymmetry breaking terms. 
Moreover, it can be substantially weakened for a non-minimal K\"ahler 
structure.         

\section{Embedding In Realistic Models}

The RH neutrino sector in Eq.~(\ref{superpot}) can be naturally added by 
extending
the MSSM to incorporate a gauged $U(1)_{B-L}$ symmetry. Three fermions, with 
the same quantum number as the RH neutrinos, will then be required for gauge
anomaly cancellation. The RH neutrinos obtain
Majorana mass through the scalar component of the $\Sigma$ superfield (with a 
$B-L$ charge of 2), which spontaneously breaks $U(1)_{B-L}$ symmetry. The 
present neutrino oscillation data
indicates the scale of symmetry breaking ${v}_{B-L}$ be somewhere 
around $10^{12}-10^{15}$ GeV. The presence of heavy RH neutrinos will ensure 
the light SM neutrino masses via the see-saw mechanism \cite{seesaw}.

Note that the inflaton is considered to be a gauge singlet, and
does not share any charge with other multiplets in
Eq.~(\ref{superpot}). Thus its coupling to the RH neutrino sector
is determined by non-renormalizable
terms which, after symmetry breaking, result in $y \sim {\cal O}
({v}_{B-L}/M_{\rm P})$.

The simplest extension of the electroweak sector has the gauge
group $SU(2)_L \times U(1)_{I_{3R}} \times U(1)_{B-L}$, with the fermion 
quantum numbers assigned
as follows: ${\bf Q}(2, 0, +\frac{1}{3})$; ${\bf L}(2, 0, -1)$; $ {\bf u}^c 
(1,-\frac{1}{2},
-\frac{1}{3})$; $ {\bf d}^c (1, +\frac{1}{2}, -\frac{1}{3})$; $ {\bf e}^c (1,
+\frac{1}{2}, +1)$; $ {\bf N} (1, -\frac{1}{2}, +1)$.
As mentioned earlier, three $N$ are required
from anomaly cancellations conditions.  The Higgs fields have the assignment 
${\bf H}_u (2,+\frac{1}{2}, 0);~ {\bf H}_d (2, -\frac{1}{2}, 0);~ \Sigma (1, 
+1, -2),~\bar{\Sigma} (1, -1, +2)$. Note that with the above charge 
assignments, two superfields $\Sigma$ 
and $\bar \Sigma$ are required for 
anomaly cancellation. The mixings and mass
differences among different neutrino flavors as observed in different
experiments can be generated in this model via flavor violating Majorana couplings~\cite{bdm}. Indeed it is possible 
to find good fits of the experimental data with Majorana
masses $> 10^{8}$ GeV~\cite{bdm}. The branching ratios of lepton flavor violating decay modes e.g. $\tau\rightarrow\mu\gamma$, $\mu\rightarrow
e\gamma$ can distinguish these models.

\section{Conclusion}

In this paper we have proposed a leptogenesis scenario where 
the lepton asymmetry is created in the RH sneutrino sector at relatively low 
reheat temperatures. This 
happens via a phase mismatch between the Majorana masses and the 
coupling of the RH (s)neutrinos to a gauge singlet inflaton. The prompt decay 
of the sneutrinos
then transfers the lepton asymmetry to the SM lepton sector. The realization 
of this scenario requires a considerable branching ratio for the inflaton 
decay to (at least one of) the RH (s)neutrinos, and new Higgs field(s) whose 
VEV is repsonsible for generating the Majorana masses. The first rquirement 
is a typical ingredient of non-thermal leptogenesis scenarios. The second one 
will be a necessary part of model building when the RH (s)neutrinos have 
gauge quantum charges under some new physics, e.g. models with a 
gauged $U(1)_{B-L}$ symmetry. The mixings and mass
differences among different neutrino flavors as observed are generated in this model 
via flavor violating Majorana couplings. The remarkable 
difference from the standard leptogenesis is that here the asymmetry 
depends on the neutrino Majorana Yukawa couplings rather the Dirac Yukawas.  
There exists another source for the wash-out of the asymmetry
in this 
scenario, in addition to the usual lepton number violating scatterings of 
leptons and Higgs, namely the scattreing of RH sneutrinos off each other or RH 
neutinos. We saw that for reheat temperatures compatible with 
the limit from thermal gravitino production, the wash-out processes do not 
lead to any meaningful constraints on the model parameters.

The maximum asymmetry is yielded when heavier (s)neutrinos have larger 
couplings to the inflaton. In this case the lepton asymmetry is 
mainly created in inflaton decay to the heaviset RH 
sneutrino ${\widetilde N}_{1}$ with mass $M_1$. An acceptable baryon 
asymmetry can then be 
obtained for moderate Majorana Yukawa couplings $g_1 \gsim 10^{-2}$, and 
$m_\phi \gsim 10 M_1$.

One important point is that the low-energy supersymmetry breaking induces the 
$\widetilde N-\bar{\widetilde N}$ oscillations and, in consequence, erases 
the initial lepton asymmetry. This demands that the decay rate of RH 
sneutrinos must be larger than the frequency of such oscillations. The latter 
quantity depends on the form of K\"ahler potential, as well as the mechanism 
for medaition of supersymmetry breaking. For minimal K\"ahler structure      
we require $M_{1} > 10^{8.5}$~GeV in 
gravity-mediated models, and $M_{1} > 10^{4.5}$~GeV in gauge-mediated 
models. These bounds can be substantially weakened for 
non-minimal cases. There will be no other constraints on $M_1$ beides this, 
and hence low scale leptogenesis can in principle be accommodated with a 
proper choice of the inflationary model.


\acknowledgements
A.M. is a CITA national 
fellow. He also acknowledges the hospitality
of the Department of Physics, University of Regina where part of this
work has been carried out. The research of R.A. and B.D. is supported by the 
National Sciences and Engineering Research Council of Canada.



\begin{references}


\bibitem{Olive:1999ij}
For a review, see: K.~A.~Olive, G.~Steigman and T.~P.~Walker,
Phys.\ Rept.\  {\bf 333}, 389 (2000)


\bibitem{Sakharov:dj}
A.~D.~Sakharov,
Pisma Zh.\ Eksp.\ Teor.\ Fiz.\  {\bf 5} (1967) 32
[JETP Lett.\  {\bf 5} (1967\ SOPUA,34,392-393.1991\ UFNAA,161,61-64.1991) 24].


\bibitem{Kuzmin:1985mm}
V.~A.~Kuzmin, V.~A.~Rubakov and M.~E.~Shaposhnikov, 
Phys.\ Lett.\ B {\bf 155}, 36 (1985).


\bibitem{Khlebnikov:sr}
S.~Y.~Khlebnikov and M.~E.~Shaposhnikov,
Nucl.\ Phys.\ B {\bf 308}, 885 (1988).

\bibitem{Fukugita:1990gb}
M. Fukugita and T. Yanagida, Phys. Lett. B {\bf 174}, 45 (1986).


\bibitem{seesaw}
M. Gell-Mann, P. Ramond and R. Slansky, in {\it Supergravity},
eds. P. van Nieuwenhuizen and D. Z. Freedman (North Holland 1979);
T. Yanagida, Proceedings of {\it Workshop on
Unified Theory and Baryon number in the Universe}, eds.
O. Sawada and A. Sugamoto (KEK 1979);
R. N. Mohapatra and G. Senjanovic, Phys. Rev. Lett. {\bf 44}, 912
(1980).



\bibitem{Luty:un}
M.~A.~Luty, Phys.\ Rev.\ D {\bf 45}, 455 (1992).

\bibitem{plumacher98}
M. Pl\"umacher, Z. Phys. C {\bf 74}, 549 (1997),
and Nucl. Phys. B {\bf 530}, 207 (1998);
W. B\"uchmuller and M. Pl\"umacher, Phys. Rept. {\bf 320}, 329 (1999),
and Int. J. Mod. Phys. A {\bf 15}, 5047 (2000).

\bibitem{Buchmuller:2002rq}
W.~Buchmuller, P.~Di Bari and M.~Plumacher, Nucl.\ Phys.\ B {\bf 643}, 
367 (2002)

\bibitem{ellis}
J. Ellis, J. E. Kim and D. V. Nanopoulos, Phys. Lett. B {\bf 145},
181 (1984); J. Ellis, D. V. Nanopoulos, K. A. Olive and S- J. Rey, Astropart. 
Phys. {\bf 4}, 371 (1996).


\bibitem{subir}
S. Sarkar, Rep. Prog. Phys. {\bf 59}, 1493 (1996).

\bibitem{maroto}
A. L. Maroto and A. Mazumdar, Phys. Rev. Lett {\bf 84}, 1655 (2000);
R. Kallosh, L. Kofman, A. D. Linde and A. Von Proeyen, Phys. Rev. D
{\bf 61}, 103503 (2000);
G.F. Giudice, I. I. Tkachev and A. Riotto, J. High Energy Phys. {\bf 9908}, 
009 
(1999);
{\bf 9911}, 036 (1999);
M. Bastero-Gil and A. Mazumdar, Phys. Rev. D {\bf 62}, 083510 (2000).

\bibitem{rouzbeh}
R. Allahverdi, M. Bastero-Gil and A. Mazumdar, Phys. Rev. D {\bf 64},
023516 (2001).

\bibitem{nps}
H. P. Nilles, M. Peloso and L. Sorbo, J. High Energy Phys. {\bf 0104}, 004 
(2001).

\bibitem{nop}
H. P. Nilles, K. A. Olive and M. Peloso, Phys. Lett. B {\bf 522}, 304
(2001).

\bibitem{allahverdi01}
R. Allahverdi, K. Enqvist and A. Mazumdar, Phys. Rev. D {\bf 65},
103519 (2002).

\bibitem{kumekawa94}
G. Lazarides and Q. Shafi, Phys. Lett. B {\bf 258}, 305 (1991);
T. Asaka, K. Hamaguchi, M. Kawasaki and T. Yanagida, Phys. Lett. B
{\bf 464}, 12 (1999),
and Phys. Rev. D {\bf 61}, 083512 (2000).

\bibitem{giudice99}
G. F. Giudice, M. Peloso, A. Riotto and I. I. Tkachev, J. High Energy
Phys. {\bf 9908}, 014 (1999).

\bibitem{enqvist02a}
K. Enqvist, S. Kasuya and A. Mazumdar, Phys. Rev. Lett. {\bf 89}, 091301
(2002), and Phys. Rev. D {\bf 66}, 043505 (2002).

\bibitem{cdo}
B. A. Campbell, S. Davidson and K. A. Olive, Nucl. Phys. B {\bf 399},
111 (1993).

\bibitem{bdps}
L. Boubekeur, S. Davidson, M. Peloso and L. Sorbo, Phys. Rev. D {\bf 67}, 
043515 (2003).

\bibitem{yanagida}
H. Murayama and T. Yanagida, Phys. Lett. B {\bf 322}, 349 (1994);
K. Hamaguchi, H. Murayama and T. Yanagida, Phys. Rev. D {\bf 65},
043512 (2002).

\bibitem{berezhiani01}
Z. Berezhiani, A. Mazumdar and A. P. Lorenzana, Phys. Lett. B
{\bf 518}, 282 (2001).

\bibitem{bento}
L. Bento and Z. Berezhiani, Phys. Rev. Lett. {\bf 87}, 231304 (2001).

\bibitem{am}
R. Allahverdi and A. Mazumdar, Phys. Rev. D {\bf 67}, 023509 (2003).

\bibitem{mrs}
E. Ma, M. Raidal and U. Sarkar, Phys. Rev. D {\bf 60}, 076005 (1999). 

\bibitem{hambye}
See, for example, T. Hambye, Nucl. Phys. B {\bf 633}, 171 (2002). 

\bibitem{one-loop}
M. Flanz, E. A. Paschos and U. Sarkar, Phys. Lett. B {\bf 345} 248
(1995); L. Covi, E. Roulet and F. Vissani, Phys. Lett. B {\bf 384},
169 (1996); A. Pilaftsis, Phys. Rev. D {\bf 56} 5431 (1997); W. B\"uchmuller 
and M. Pl\"umacher, Phys. Lett. B {\bf 431}, 354 (1998).

\bibitem{pilaftsis}
A. Pilaftsis, Int. J. Mod. Phys. A {\bf 14}, 1811 (1999).


\bibitem{thermal} 
S. Davidson and S. Sarkar, J. High Energy Phys. {\bf 0011}, 012 (2000); 
R. Allahverdi and M. Drees, Phys. Rev. D {\bf 66}, 063513 (2002);  
P. Jaikumar and A. Mazumdar, hep-ph/0212265.

\bibitem{fy}
M.~Fukugita and T.~Yanagida, Phys. Rev. D {\bf 42}, 1285 (1990).

\bibitem{bdm}
K.~S.~Babu, B.~Dutta and R.~N.~Mohapatra, hep-ph/0211068, Phys. Rev. D 
{\bf 67}, 076006 (2003).




\end{references}
\end{document}